\documentstyle[preprint,aps]{revtex}
\begin{document}
\draft
\title{Experimental evidence for fast cluster formation of chain oxygen
vacancies in $\bf YBa_2Cu_3O_{7-\delta}$ being at the origin of the
fishtail anomaly} 

\author{Andreas Erb, Alfred A. Manuel, Marc Dhalle, Frank Marti,
Jean-Yves Genoud\cite{JeanYves}, Bernard Revaz, Alain Junod, Dharmavaram
Vasumathi, Shoji Ishibashi\cite{Shoji}, Abhay Shukla\cite{Abhay},
Eric Walker, \O ystein Fischer and Ren\'e Fl\" ukiger}

\address{ D\'epartement de Physique de la Mati\^^ ere Condens\'ee,
Universit\'e de Gen\^^ eve, 24, quai Ernest Ansermet, CH-1211 
Gen\^^ eve 4, Switzerland}

\author{Riccardo Pozzi, Mihael Mali and Detlef Brinkmann}

\address{ University of Z\" urich, Physics Department, Winterthurerstr.
190,CH-8057 Z\" urich Switzerland}

\date{Version 1.3, \today}
\maketitle

\begin{abstract} 
We report on three different and complementary measurements, namely
magnetisation measurements positron annihilation spectroscopy and NMR
measurements, which give evidence that the formation of oxygen vacancy
clusters is on the origin of the fishtail anomaly in $\rm
YBa_2Cu_3O_{7-\delta}$. While in the case of $\rm YBa_2Cu_3O_{7.0}$ the
anomaly is intrinsically absent, it can be suppressed in the optimally
doped state where vacancies are present. We therefore conclude that the
single vacancies or point defects can not be responsible for this
anomaly but that clusters of oxygen vacancies are on its origin.
\end{abstract}

\narrowtext

Soon after the discovery to the 90 K superconductor $\rm
YBa_2Cu_3O_{7-\delta}$ an anomaly in the curves of the irreversible
magnetisation, often referred to as {\it fishtail effect}, was reported
\cite{daumling}. This anomaly consists of an increase in the
magnetisation upon increasing magnetic field which is equivalent to an
increase of the critical current high above $\rm Hc_1$. Whereas a technical
use of all other high temperature superconductors at liquid nitrogen
temperatures is actually limited due to their poor critical current
density in magnetic fields, the fishtail anomaly of the 123
superconductors makes them potentially useful in technical applications
even at high fields. The origin of this anomaly, however, was
controversially discussed for nearly a decade. Recently, we reported
how the fishtail anomaly can be reversibly suppressed in the optimally
doped state of $\rm YBa_2Cu_3O_{6.92}$ and showed that it is completely
absent in the fully oxygenated state $\rm YBa_2Cu_3O_{7}$ \cite{erb1}.
In the present study, we give independent proofs for the formation of
oxygen vacancy clusters being responsible for the effect. Understanding
the formation mechanism of such clusters allows us to tune the pinning
properties of $\rm YBa_2Cu_3O_{7-\delta}$ to either low pinning for fundamental studies such
as the investigation of vortex matter or to strong pinning by
deliberately clustering the vacancies to a desired size, thus changing
the critical current density and the irreversibility field.

The experiments were performed on crystals which were grown in the
recently developed non- reactive crucible material $\rm BaZrO_3$
\cite{erb2}. Crystals obtained from such experiments exhibit a superior
purity since they are not contaminated with metallic impurities from
the container material \cite{erb3}. Wet chemical analysis (ICP-MS) on
three of the $\rm YBa_2Cu_3O_{7-\delta}$ crystals showed that Zr, Sr
and La with concentrations between 0.0005 (detection limit) and 0.001
at. \%, are the only impurities. The crystals used for the oxygenation
experiments exhibited the usual twinning within the a-b plane and no
attempts were made to detwin the crystals. The standard annealing to
obtain high transition temperatures well above 90 K in $\rm
YBa_2Cu_3O_{7-\delta}$ single crystals, consists of a treatment under 1
bar of oxygen at around 500 $\rm ^oC$ for 100 - 200 hours, followed by a quench
to room temperature. However, the oxygenation of $\rm YBa_2Cu_3O_{7-\delta}$ is performed in
this way only due to the simplicity of the procedure. According to the
phase diagram of Lindemer et al. \cite{lindemer} which gives the
equilibrium oxygen concentrations as a function of oxygen partial
pressure and temperature, the same oxygen content can be obtained as
well by annealing at higher temperatures in higher pressure oxygen
atmospheres. The advantage of annealing at higher temperatures and
higher oxygen partial pressures is that it results in an oxygenation
which is presumably more homogenous and in shorter annealing times. We
performed high pressure oxygen treatments at 650 - 750 $\rm ^oC$ in 100 bar of
$\rm {O_2}$ for periods of 12 - 40 h, followed by rapid quenching (about 1 min)
to room temperature, so that the high temperature oxygen state can be
maintained. For the case of optimally doped $\rm YBa_2Cu_3O_{7-\delta}$ we have chosen
combinations of oxygen partial pressures and temperatures which lead to
the same equilibrium oxygen content in the crystal.

The magnetisation measurements were performed in a SQUID Magnetometer
(Quantum Design) with the magnetic field parallel to the c-axis of the
crystal. The result of an oxygenation treatment at 700 $\rm ^oC$ for 12.5 h at
100 bar can be seen in Fig. 1, where M(H) is shown for a temperature of
70 K.

No fishtail behaviour is present after this oxygenation treatment.
Re-annealing the same crystal similar to the standard oxygenation (510
$\rm ^oC$, under 1 bar oxygen) leads to the re-appearance of the fishtail
anomaly, and this already after a short re-annealing time of only half
an hour. If the crystal is further annealed at this temperature, the
maximum in the reversible magnetisation, which was initially been
re-established at a value of around 3.3 T, shifts to lower field values
of about 1.5 T for an annealing time of 160 hours, whilst the amplitude
of the maximum increases indicating a increase in the critical current.
Note that, according to \cite{lindemer}, the overall oxygen content of
the sample has the same value of 6.92 for the three treatments and that the
transition temperatures are not changed.

This time evolution of the anomaly is a direct confirmation of our
earlier interpretation \cite{erb1,erb3} of clusters of oxygen vacancies
being responsible for the fishtail effect. At temperatures as high as
700 $\rm ^oC$ the oxygen deficient regions are more likely to be randomly
distributed, while at lower temperatures oxygen deficient unit cells
tend to cluster due to the smaller contribution of the entropy term to
the free energy \cite[and references therein]{erb1,erb3,vargas}. Thus,
when we re-anneal the crystals clusters form. Their size and mean
distance however depend on the re-annealing time. Short time annealing
produces small and densely spaced clusters, which grow, when annealing
progresses, to bigger size with larger distances. Pinning of vortices
is most effective if the distance between vortices matches to the mean
distance between the pinning centers. In the case of pinning by oxygen
vacancy clusters, we find, for the short re-annealing time of half an
hour a maximum for the critical current at higher fields (3.3 Tesla),
which is shifted to lower fields for longer annealing times, when
cluster growth makes their mean distance larger. On the other hand the
amplitude of the maximum increases for the longer anneals since the
bigger clusters have a higher pinning potential.

A simple picture defining the maximum of the anomaly as a matching
field enables us to calculate the average spacing between oxygen
vacancy clusters. Doing so we find a spacing of about 25 nm for a
matching field of 3.3 T and 37 nm for the maximum at 1.5 T
respectively. Assuming that all vacancies present within the space
between clusters would contribute to its formation and that the
clusters consist of $\rm YBa_2Cu_3O_{6.0}$, one calculates for the
present concentration of $\rm YBa_2Cu_3O_{6.92}$ that around 350
vacancies are in a clustered state after the short term annealing. Thus
an upper limit for the size of a cluster is about 18 x18 unit cells.
For longer annealing these clusters would then contain some 750
vacancies, setting the upper limit for cluster sizes to 27 x 27 unit
cells. We note here that certainly not all vacancies are in a clustered
state since $\rm {T_c}$ remains unchanged, so that the estimation above only
gives an upper
limit for the size. 

That such a short range reorganisation of the vacancies is indeed
possible can be seen by calculating the effective diffusion length,
using the diffusion coefficients measured on identical samples
\cite{erb4,klaser}. The chemical diffusion coefficient, which is the
product of the self-diffusion coefficient and the thermodynamic factor,
turned out to be $\rm 8 x 10^{-8} cm^2/s$ at 510 $\rm ^oC$ with the thermodynamic factor
being about 10 \cite{conder,faupel}. Using the Einstein formula 
$d_{eff} = \sqrt{6D_{self}(T)t}$, an
estimation of the effective diffusion length for the short re-annealing
of half an hour yields an possible migration of vacancies of about 90
mm, which is far more than needed. 

Among the methods, which allow us to study the oxygen vacancies more
directly, are the positron annihilation technique and NMR which we will
discuss now. Both methods probe bulk properties and this on a
microscopic level. In addition, these techniques do not alter the
oxygen vacancy distribution.

Positron annihilation spectroscopy \cite{dupasquier} is site-sensitive
and probes bulk properties. Measured spectra depend strongly on the
chemical and structural environment of the annihilating positron. In
$\rm YBa_2Cu_3O_{7-\delta}$, it has been established that the positrons
i) annihilate in the region of the copper-oxygen chains \cite{stetten};
ii) tend to localize near oxygen vacancies which function as weak {\it
traps} \cite{peter}; and iii) are trapped more efficiently by larger
oxygen vacancy clusters \cite{ishibashi}.

We have measured the quantity $N(p_x,p_y)$, the electron momentum
density sampled by the annihilating positrons. The temperature of the
twinned $\rm YBa_2Cu_3O_{7-\delta}$ single crystal was 177 $\rm ^oC$ to
favour the thermally activated migration of the positrons trapped by
shallow structural defects like single oxygen vacancies \cite{peter}.
The momentum-dependent annihilation lines $N(p_x,0)$ are shown in Fig.
2 for the three different sample treatments: a) fully oxygenated
(triangles), b) optimally doped without fishtail (squares) and c)
optimally doped with fishtail (circles). The $N(p_x,p_y)$ distributions
were normalised to equal area. The peak height of $N(p_x,0)$, in a
sample containing defects acting as positron traps, corresponds to the
degree of positron localization.  Trapped positrons are more likely to
annihilate with valence electrons than with bound core electrons of
higher momenta (see \cite{dupasquier}), giving rise to a line more
peaked at the zero momentum. We observe the lowest peak (indicating
delocalized positrons) in the case of the fully oxygenated $\rm
YBa_2Cu_3O_{7-\delta}$, as there is no trapping by oxygen vacancies. In
the optimally doped state with fishtail, the annihilation line has the
highest peak. This signifies the presence of large clusters of oxygen
vacancies which act as more efficient positron traps than single oxygen
vacancies or small clusters \cite{peter}. In the optimally doped state
without fishtail, the peak height is somewhat larger than for the fully
oxygenated state, indicating weak positron trapping, which is
attributed to residual trapping by clusters too small to act as pinning
centres for vortices. Our interpretation is further confirmed by the
difference of peak height previously observed between $\rm
YBa_2Cu_3O_6$ (insulating) and $\rm YBa_2Cu_3O_7$ crystals
\cite{barbiell}. $\rm YBa_2Cu_3O_6$ shows the highest peak height,
similar to what we observe for the optimally doped state with fishtail.
This is in agreement with the fact that, in a large oxygen vacancy
cluster, the positron sees a local environment corresponding to the
$\rm YBa_2Cu_3O_6$ phase.

The correlation between the large peak height in $N(p_x,0)$ and the
presence of the fishtail leads us to conclude that the fishtail anomaly
in the magnetisation is a peak effect induced by the clustering of the
oxygen vacancies.

Finally, we will discuss our NMR studies which probe microscopically
the arrangement of the oxygen vacancies. We employed Cu NMR which is an
appropriate method since the resonance frequency of Cu1 (chain copper)
depends strongly on the co-ordination by nearest-neighbour oxygen
ions.  The magnetic shift and the electric field gradient tensors are
well-known for plane copper, Cu2 \cite{pennington}, copper in a filled
chain, $\rm {(Cu1)_4}$, i.e. co-ordinated by four oxygen ions
\cite{pennington} and copper in an empty chain, $\rm {(Cu1)_2}$, i.e.
co-ordinated by two apex oxygen only \cite{mali}. We investigated a
single crystal whose magnetization curves, for two oxygenation states,
are shown in Fig. 3.  For the $\rm YBa_2Cu_3O_{7}$ state, NMR yielded
four Cu peaks (Fig. 4a):  Two of them are very narrow and correspond to
the Cu2 and $\rm {(Cu1)_4}$ central lines ($\rm {-1/2 \leftrightarrow
+1/2}$ transitions) of the quadrupolar splitting and the broader double
peak at 102.15 MHz consists of the two satellite lines ($\rm {\pm 3/2
\leftrightarrow \pm 1/2}$ transitions) of $\rm {(Cu1)_4}$. The
intensity of a line is given by the integral over the line and is
proportional to the number of nuclear spins contributing to the
corresponding line. As expected by virtue of the crystal structure, the
intensity of the Cu2 central line turned out to be roughly twice the
intensity of the $\rm {(Cu1)_4}$ central line. The absence of other Cu
lines indicates that the chains are perfectly filled and thus confirms
the extremely low defect concentration of the fully oxygenated crystal.
In addition, a tiny signal of sodium showed up at 101.3 MHz, probably
stemming from sweat in the crumpled piece of cotton wool fixing the
crystal.

Next, we investigated the $\rm YBa_2Cu_3O_{6.9}$ state. If the crystal
is annealed at 510 $\rm ^oC$, the lines discussed above become much
broader (Fig. 4b). Furthermore, a new signal appears at 101.4 MHz that
corresponds to chain $\rm {(Cu1)_2}$. By comparing the Cu2 and $\rm
{(Cu1)_2}$ line intensities, corrected for their spin-spin relaxation
times and frequencies, we conclude that 5 - 10\% of all chain Cu ions
are co-ordinated by an oxygen vacancy on either side. Furthermore, the
very narrow $\rm {(Cu1)_2}$ line (with a "full width half height" of 13
kHz) reveals a high degree of local order at this site, indicating that
$\rm {(Cu1)_2}$ must be located in rather large empty-chain clusters.
The formation of such large clusters would also explain why we did not
detect $\rm {(Cu1)_3}$ signals which would arise from Cu sites with
only one oxygen vacancy neighbour in the chain. The $\rm {(Cu1)_3}$
site appears at the interface between filled and empty chain segments
with a relative abundance that diminishes with increasing length of the
empty segments. The corresponding NMR experiments with $\rm
YBa_2Cu_3O_{6.9}$ and suppressed fishtail are in progress. However,
they are very time consuming since the sample size must be kept
sufficiently small to allow a high quenching rate which is necessary
for a complete suppression of the anomaly.

In conclusion three different and complementary measurements give
evidence that the formation of oxygen vacancy clusters is the origin of
the fishtail anomaly in $\rm YBa_2Cu_3O_{7-\delta}$. Note, that the
single vacancies or point defects can not be responsible for this
anomaly, since this anomaly can also be suppressed in the optimally
doped state where vacancies are present. Thus, a formation mechanism
that leads to larger nevertheless still microstructural inhomogeneities
must be responsible for this effect. In the case of $\rm
YBa_2Cu_3O_{7.0}$ the anomaly is intrinsically absent unless the
crystals contain metallic impurities. On such $\rm YBa_2Cu_3O_{7.0}$
samples melting of the vortex lattice has been observed by calorimetric
measurements \cite{junod} as a first order transition between 4 and
26.5 T, which was the highest available field, giving further support
for our interpretation and yields new insight into the nature of vortex
matter. For application purposes the understanding of the formation
mechanism of microstructural inhomogeneities offers the possibility to
tailor the samples according to the required properties influencing the
maximum value of the critical current and the irreversibility field.

\begin{figure}
\caption{Magnetisation of $\rm YBa_2Cu_3O_{6.92}$ at 70 K after
different annealing regimes. The fishtail anomaly is re-established
already after short annealing times and moves to lower field values for
longer anneals.}
\label{fig1}
\end{figure}

\begin{figure}
\caption{Momentum-dependent positron annihilation lines for $\rm
YBa_2Cu_3O_{6.9}$ with fishtail (circles), without fishtail (squares)
and for $\rm YBa_2Cu_3O_{7}$ (triangles).}
\label{fig2}
\end{figure}

\begin{figure}
\caption{Magnetisation curves at 70 K of $\rm YBa_2Cu_3O_{x}$ - single
crystal AE276G after full oxygenation to a $\rm O_7$ (empty symbols) and
$\rm O_{6.90}$ (full symbols).}
\label{fig3}
\end{figure}

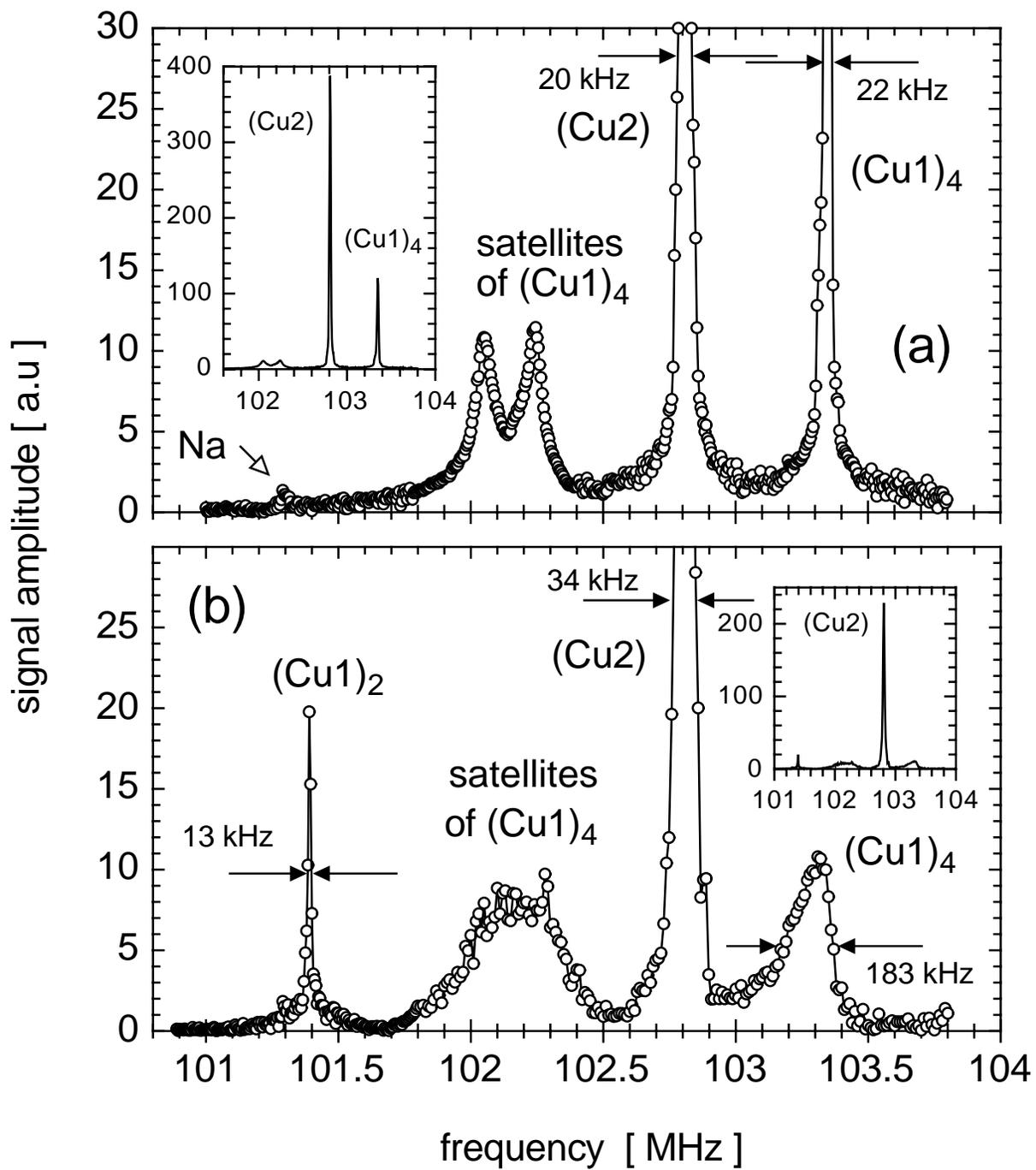
\begin{figure}
\caption{$\rm ^{63}Cu$ central (and satellite) lines of different sites are
shown (a) for $\rm YBa_2Cu_3O_{7}$ and (b) for $\rm YBa_2Cu_3O_{6.9}$
at room temperature. The external magnetic field (8.995 Tesla) was
oriented along the c-axis of the single crystal. The indicated
linewidths are defined as the full widths at half maximum. The insets
show the full size of the lines.}
\label{fig4}
\end{figure}

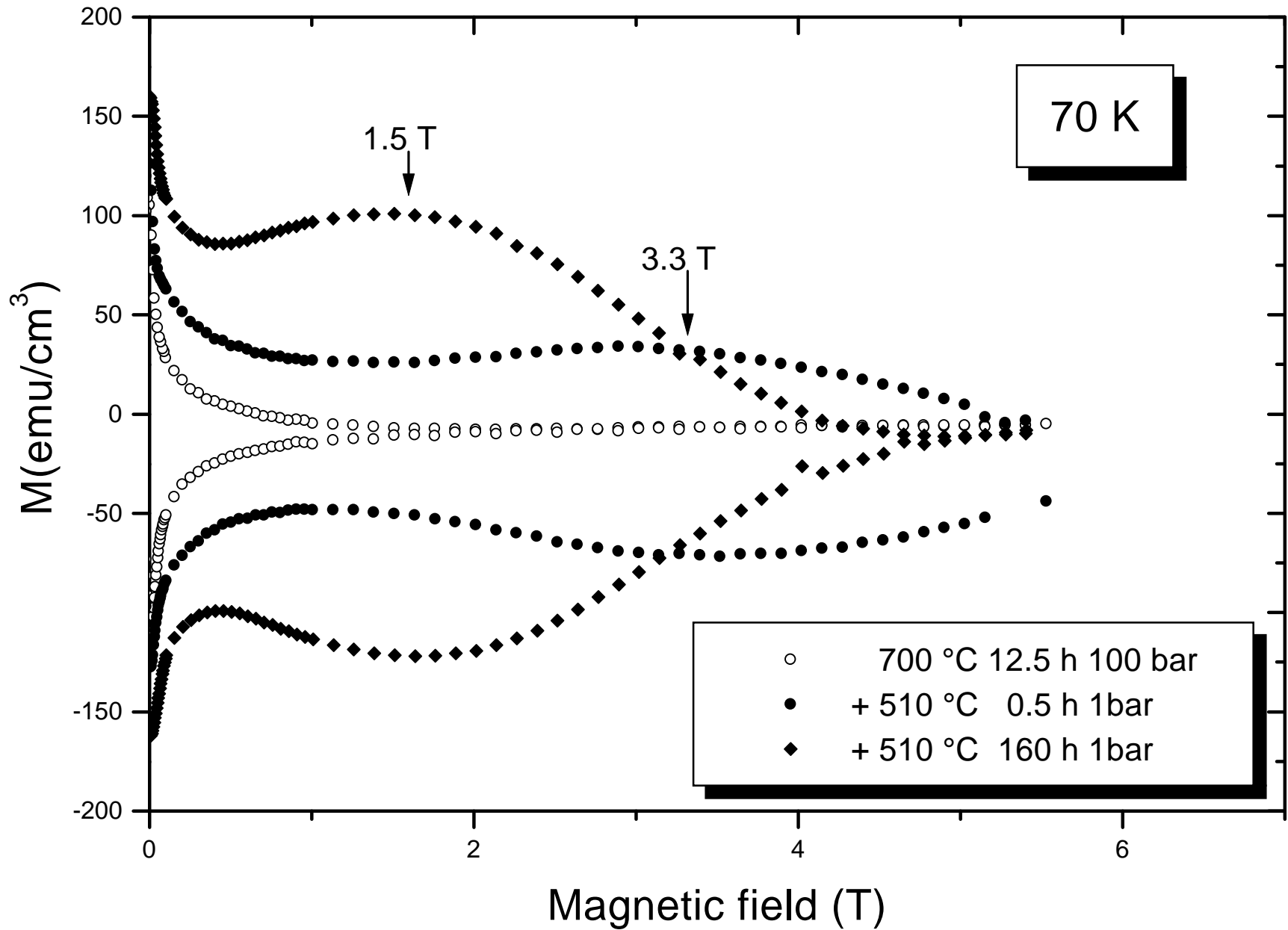
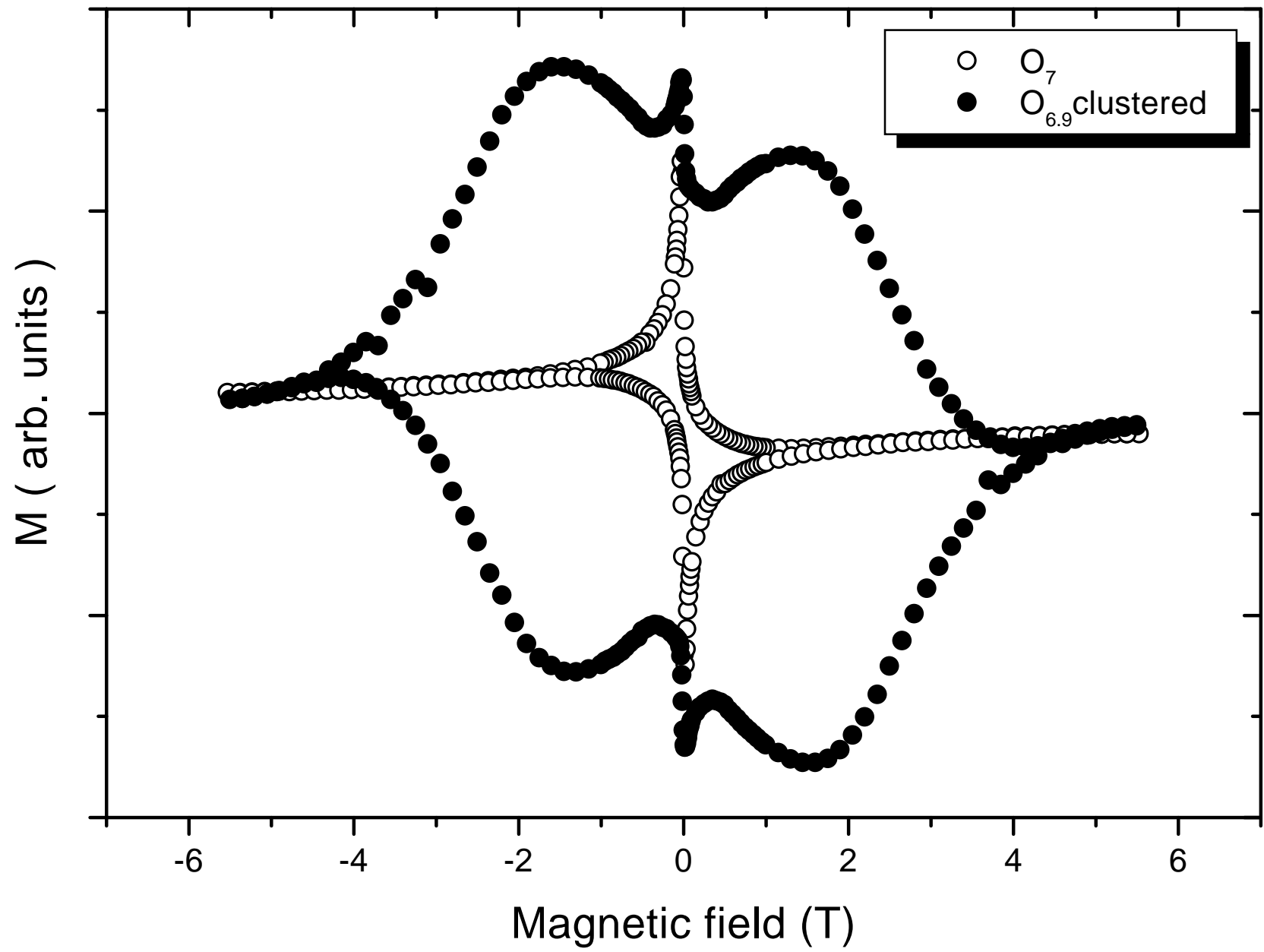
\end{document}